\documentclass[12pt]{article}
\usepackage{latexsym,amsmath,amssymb,amsbsy,graphicx}
\usepackage[utf8]{inputenc}
\usepackage[T2A]{fontenc}
\usepackage[space]{cite} 

\makeatletter\def\@biblabel#1{\hfill#1.}\makeatother

\allowdisplaybreaks
\begin {document}

\noindent\begin{minipage}{\textwidth}
\begin{center}

{\Large{Spectral energy distribution of late stage stars}}\\[9pt]

{\large Tatarnikov A.\,M.$^{1,2a}$, Zheltoukhov S.\,G.$^{1,2}$, Malik E.\,D. $^{1,3}$}\\[6pt]

\textit {$^1$Faculty of Physics, M.V.Lomonosov Moscow State University, Moscow 119991, Russia.}\\
\textit {$^2$Sternberg Astronomical Institute,
M.V.Lomonosov Moscow State University, Moscow 119234, Russia.}\\
\textit {$^3$Institute of astronomy of Russian Academy of Sciences, Moscow 119017, Russia.}\\
\textit {E-mail: $^a$andrew@sai.msu.ru}\\

\end{center}

{\parindent5mm This paper presents a catalog of the energy distribution in the spectra of 263 stars in the wavelength range from 0.4 to 100~$\mu$m, which are at late stages of evolution and have been observed by the ISO space observatory. For each object in the catalog, estimates of the observed bolometric fluxes were derived from smoothed energy distribution curves. The catalog is available at \textit{https://infra.sai.msu.ru/sai\_lss\_sed} both as a table and in machine-readable format. It is shown that for the specified sample of objects their ISO SWS spectra in the range 2.4~- 45~$\mu$m only 60\% of cases correspond to the general shape of the continuum, and can be used without recalibration. A selection of carbon stars, accessible for the infrared observations from the MSU observatories has been made. For some of them the first brightness estimates in the $K, L$, and $M$ bands were obtained with the new IR camera of the 2.5-m telescope of CMO.

\vspace{2pt}\par}
\vspace{1pt}\par

\vspace{1pt}\par
\end{minipage}



\section*{Introduction}
\mbox{}\vspace{-\baselineskip}

Stars in the AGB and post-AGB stages are important suppliers of matter to the interstellar medium. These stages are the late stages of evolution to which stars with an initial mass of $0.5-8 M\odot$ fall (\cite{Iben1983}, \cite{Hofner2018}). The mass loss rate during these stages can range widely from $10^{-8} M_\odot$/yr to $10^{-5} M_\odot$/yr and in some cases up to $10^{-4} M_\odot$/yr. The ejected medium moves away from the star at the velocity of $\sim10-15$~km/s, forming an extended gas-dust envelope around the star. Depending on the ratio of the C/O content in the stellar atmosphere the composition of dust grains in the envelope will be different. Thus, at C/O>1 the star becomes a so-called carbon star with absorption bands of C$_2$, HCN, C$_2$H$_2$, etc. in its spectrum, and carbon and silicon carbide dust particles will predominate among the dust, with a typical emission feature at a wavelength of 11.3~$\mu$m \cite{SiC}.

The study of circumstellar dust envelopes is practically impossible without infrared observations. This is due to the typical temperatures of both stars at the AGB stage ($T_*\le3500$~K) and dust in the envelopes ($T_{dust}<1500$~K~--- their emission maxima lie in the IR range. The wider the spectral range in which the data are obtained, the more reliably the parameters of the envelope and the star are determined, from the chemical composition of the dust grains to their size, the law of distribution of matter in the envelope, and, finally, the mass and size of the envelope.

In the mid-to-late 1990s, the ISO \cite{Kessler1996} (Infrared Space Observatory) operated in Earth orbit. It carried several instruments, notably the Short Wavelength Spectrometer (SWS, \cite{deGra1996}), which operated over a wide spectral range from 2.36 to 45~$\mu$m. One of the main classes of objects observed with this instrument were AGB stars of various types and post-AGB stars. Among the nearly 900 objects observed with SWS, at least 263 belong to these types of stars.

ISO spectra alone are not sufficient to make an informative spectral energy distribution (SED) of a dust envelope and a star ~--- observations at shorter wavelengths provide information about the stellar component, and at longer wavelengths~--- about the outer, colder regions of the envelope, where the bulk of the envelope may be concentrated. The goal of our work is to create a catalog of SEDs in the widest spectral range for late-evolutionary stars observed during the ISO mission, and then select objects for a program to study the circumstellar dust envelopes of carbon stars.

\section{Observations}
\mbox{}\vspace{-\baselineskip}

The observations in the bands $K$, $L$, and $M$ were carried out with the mid-infrared camera LMP (\cite{Zheltoukhov2022}, \cite{Zheltoukhov2024}) mounted on the 2.5-m telescope of the Caucasian Mountain Observatory of SAI MSU (CMO, \cite{Shatsky2020}). The stars from Table~6 of the \cite{Shenavrin2011} paper, which were at close air mass to each of the observed objects, were used as comparison stars. Modulation of the light flux was performed by moving the telescope every 5 seconds by an angle of $10''$. Brightness measurements were made according to the scheme "object1~-- standard1~-- object2~-- standard2, etc". The time between observations of the object and the corresponding standard did not exceed 7 minutes. The results of the observations are presented below in Table~\ref{table:phot}.

\section{Entry data}
\mbox{}\vspace{-\baselineskip}

We used the list ''An atlas of fully processed spectra from the SWS''\footnote{https://users.physics.unc.edu/~gcsloan/library/swsatlas/atlas.html} (hereafter Atlas) as a source of flux data in ISO spectra. The spectra processing and normalization methods used to compile this Atlas are described in \cite{Sloan2003}. For each entry in the Atlas, the object name, unique ISO observation number (TDT~-- target dedicated time), equatorial coordinates, notes, and energy distribution tables are given. The Atlas contains 1248 records with spectra of about 900 objects obtained in the AOT1 imaging mode, which involves observations in the wavelength range from 2.36 to 45~$\mu$m, as well as 23 spectra obtained in the higher resolution and higher sensitivity AOT6 mode.

Firstly we selected stars at late stages of evolution from the Atlas. For this purpose the coordinates of the object were queried to the Simbad Astronomical Database - CDS (Strasbourg)\footnote{http://simbad.u-strasbg.fr/simbad/} and the list of types with which the object is labeled in this database was taken as a response. A total of 263 objects with types AGB, Post-AGB, PPN and PN, Long Period Variable (LPV), Mira, Carbon star and Red Supergiant were selected.

When studying objects with circumstellar dust shells, it is important to analyze the SED in the widest spectral range, which should include both the maximum emission of the central source and the long-wavelength emission of the dust shell. This facilitates SED modeling and allows us to obtain a reliable estimate of the bolometric flux (and at a known distance and under the assumption of spherical symmetry of the shell or when observing a disk-shaped shell "flat"{}~--- the luminosity of the central star). The long-wavelength limit of the range is usually determined by the availability of measurements of the object in all-sky IR surveys (e.g., IRAS) and should be at least 100~$\mu$m.

The short-wavelength boundary is determined by the position of the emission maximum of the central star. In most cases, for objects at late stages of stellar evolution, the energy source is a cold star with $T_{\rm eff} < 4000$\,K (exceptions are proto-planetary nebulae and relatively hot post-AGB objects). In this case, with a sufficiently dense dust envelope, on the short-wavelength side it is sufficient to have brightness estimates in the $R, I$ or $r, i, z$ bands. For miras and other cold objects with such a shell and large variability in the visible wavelength range (several stellar magnitudes in the $V$ band), it is sufficient to have data starting from band $J$ of the 2MASS survey. This facilitates the integration of data obtained at different phases of the variability since the amplitude of such objects in the near-IR is small and amounts to $\approx 1^m$ \cite{Fedoteva2020}, \cite{Tatarnikov2024}. Note that in the calculations of radiative transfer and temperature balance in the shell, the short-wavelength range boundary should be smaller than the data available in the SED. Otherwise, the magnitude of the radiation absorbed by the envelope may be underestimated. For example, for a star of spectral type M0III, neglecting radiation at $0.3 < \lambda < 0.7$~$\mu$m will result in a loss of 15\% of energy, and for M5III~--- 3\%.

We collected information on the photometry data and ISO SWS spectra using the sedbys code \cite{sedbys}, which was created to compile the SEDs of young stars. It uses as common catalogs 2MASS \cite{2mass}, AKARI \cite{akari}, GALEX \cite{galex}, Gaia \cite{gaia}, Tycho-2 \cite{tycho2}, IRAS \cite{iras}, JCMT \cite{jcmt}, APASS \cite{apass}, MSX6C \cite{msx6c}, SDSS \cite{sdss}, SPITZER \cite{spitzer}, WISE \cite{wise}, XMMOM \cite{xmmom}, and catalogs of observations of young objects. We have eliminated the reference to the last, and supplemented the code with a reference to the ISO LWS spectra catalog (data from the long-wavelength spectrograph operating in the 43~-- 197~$\mu$m range) and a conversion of the data to unified units. The observational data collected in this way form the basis of the SED catalog of the stars in the late stages of evolution.

\section{Calculation of bolometric fluxes}
\mbox{}\vspace{-\baselineskip}

\begin{figure}[h]
\center{\includegraphics[width=0.9\linewidth]{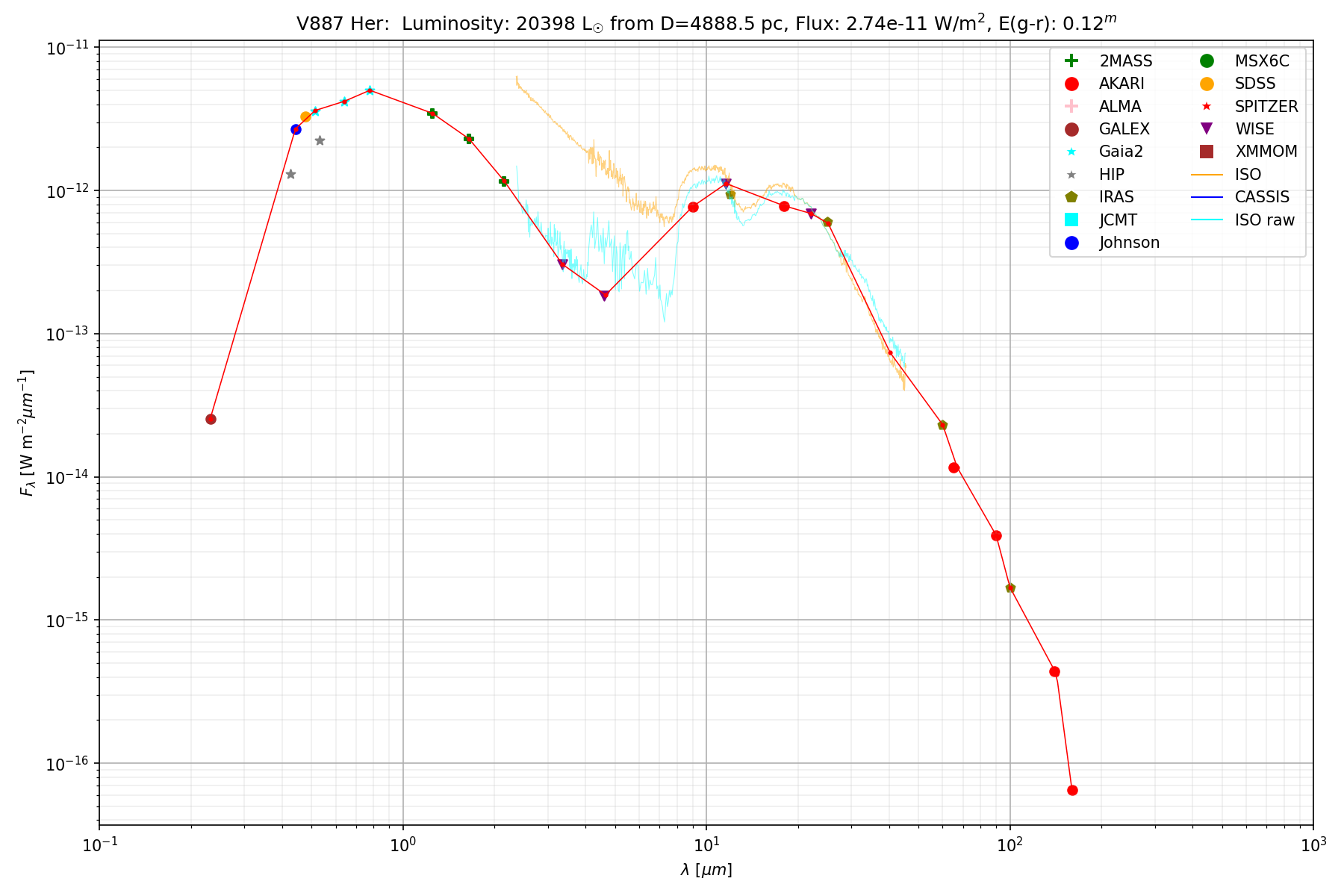}}
\caption{An example of an incomplete match between SED of the known post-AGB star V887~Her from photometric observations (symbols in different colors) and the ISO SWS spectrum from the Atlas (orange line). The red line shows the smoothed SED, from which the bolometric flux was calculated}
\label{fig:fig1}
\end{figure}

A comparison of the ISO SWS spectral data from the Atlas and the SEDs constructed from photometric data shows that in a number of cases they do not coincide. An example of a typical SED of such an object is shown in Fig.~\ref{fig:fig1}. It can be seen that in the long-wavelength range the ISO spectrum matches the photometric SED, while in the short-wavelength region (2.4~-- $\sim$7~$\mu$m) the fluxes in the spectrum are several times larger. Such large differences cannot be explained by the variability of this object \cite{arkhipova1993}, and, therefore, they must be related to the unsuccessful calibration of the spectrum in Atlas. This is also indicated by the better match of the ISO spectrum, which was downloaded directly from the ISO website (blue line in Fig.~\ref{fig:fig1}) and smoothed with a median filter to reduce noise.

We found that only less than 60\% of the ISO spectra from the Atlas can be used over the full range without recalibration. Another 20\% require relatively minor changes and can be used in the work, and the remaining spectra cannot be used in the study of SED objects without new processing of the original data.

\begin{figure}[h]
\center{\includegraphics[width=0.9\linewidth]{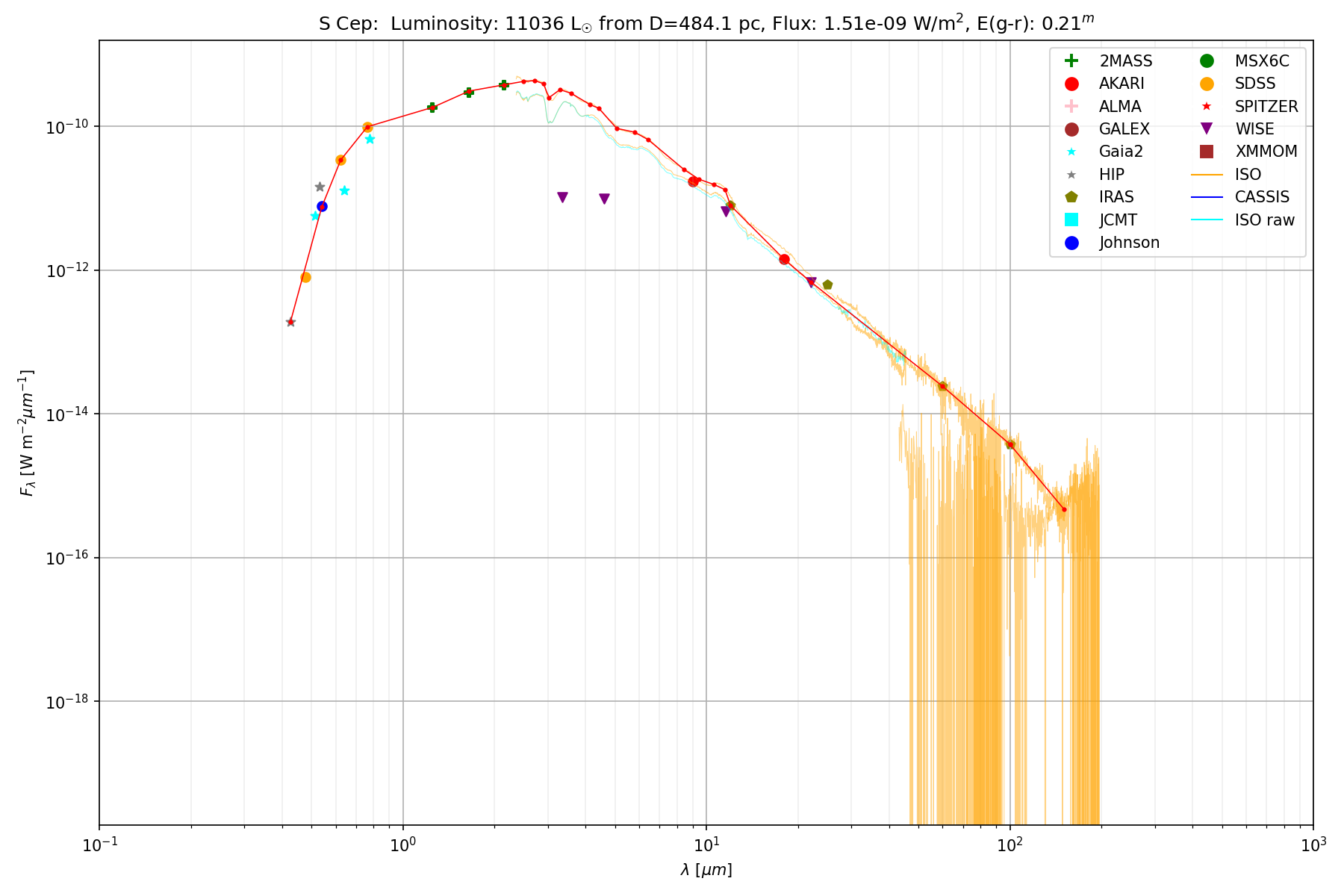}}
\caption{An example of a good match between SED of the bright carbon star S~Cep and the ISO SWS spectrum. The designations are the same as in Fig.~\ref{fig:fig1}}
\label{fig:fig2}
\end{figure}

Fig.~\ref{fig:fig2} shows an example of a good agreement between the ISO SWS spectra obtained at two pulsation phases of the carbon star S~Cep and the photometric SED. We note the significant deviation of the fluxes in the short-wavelength bands $W1$ ($\lambda=3.35$~$\mu$m) and $W2$ ($\lambda=4.6$~$\mu$m) of the WISE survey from the total SED line. This is likely due to the detector overexposure by the emission of such a bright object as S~Cep ($K \approx 0^m$, $L \approx M \approx -1^m$). There are many bright stars among the objects observed by ISO, and this behavior of the WISE flux estimates (at flux magnitudes $>5 \cdot 10^{-12}$~W/m$^2$\,$\mu$m) should be taken into account in the SED analysis.

Another factor complicating the automatic compilation of SED and the calculation of the bolometric flux from it, is the brightness variability mentioned above, which is characteristic of most stars at late stages of evolution. In the visible range, this variability can lead to a flux change of 1-2 orders of magnitude, in the near-IR~--- by a factor of $\sim 2$, and in the mid- and far-IR~--- up to 20-30\% (see Fig.~\ref{fig:fig2}). At the same time, due to the fact that the emission maximum of most objects in our sample lies in the near-IR range, the variability in the near-IR approximately corresponds to the variability of the bolometric flux, while the variability in the visible spectrum changes the bolometric flux only by a few percent.

Because of the factors described above, the final SED, from which the bolometric flux was calculated, was performed on the collected points manually. To assist in this, a program was written that divided the entire spectral range into 20 bins evenly spaced on a logarithmic wavelength scale, and in each range selected the photometric observations with the highest flux (the spectral data were not analyzed by the program). The SED obtained by the program was shown to the user, who used the mouse pointer to delete or rearrange the points suggested by the program and add new ones where necessary, including those based on the ISO and CASSIS spectra.

To calculate the bolometric flux, linear interpolation was performed between manually (using the program) placed SED points in the logarithmic wavelength scale (an example of such interpolation is shown by the red line in Figs.~\ref{fig:fig1} and \ref{fig:fig2}). It is quite difficult to obtain an estimate of the accuracy of the bolometric flux determination due to some subjectivity in carrying out the continuum level when constructing the SED. A comparison of the results obtained for well-studied stars, such as V~CrB \cite{Fedoteva2020} and T~Dra \cite{Tatarnikov2024}, shows that in the case of sufficiently detailed SEDs and with a good match between the ISO spectrum and the photometric data, the bolometric flux errors probably do not exceed 20\%. Accuracy for stars whose SEDs are poorly described by the available observational data, such as V4334~Sgr or LP~And, is expected to be lower.

To calculate the luminosity, we used the distance returned on request from the SIMBAD Astronomical Database and accounted for interstellar absorption according to the absorption map \cite{Green_2019} and the law of interstellar reddening \cite{Cardelli1989}. In most cases, SIMBAD gave data from Gaia EDR3 \cite{gaia_edr3}. In the absence of distance data, the luminosity was calculated for a distance of 1~kpc and in the absence of interstellar absorption.

\mbox{}\vspace{-\baselineskip}
\section{Description of the SED catalog of stars in late stages of evolution}
\mbox{}\vspace{-\baselineskip}

The catalog of SED stars observed by the ISO and in the late stages of evolution, which we have collected, is available on the website of the Infrared Astronomy Group of the SAI MSU: \textit{https://infra.sai.msu.ru/sai\_lss\_sed}. It contains 263 objects: 63 miras and AGB stars, 56 carbon stars, 48 post-AGB stars, and 96 objects of other types (S stars, protoplanetary nebulae, long-period variables, red supergiants, etc.). It should be noted that the SIMBAD classification was used when counting stars of different types and when indicating the type in the catalog. For example, the General Catalog of Variable Stars \cite{Samus2017} classifies some carbon stars as miras and some AGB stars as semi-regular variables.

\begin{table}[t]
\caption{Description of SED catalog columns}
\begin{center}
\scriptsize
\begin{tabular}{| l | l | l | l |}
\hline
№ & column & Description & Notes \\
\hline
1 & Object & Object name & In SIMBAD readable form \\
2 & Type & Object type & Object type and spectral type  \\
  &      & (according to SIMBAD) &  (if available) \\
3 & R.A. & Right Ascension, & On the equinox J2000 \\
  &      & HH:MM:SS.ss & \\
4 & Dec & Declination, DD:MM:SS.ss & On the equinox J2000  \\
5 & J mag & $J$ brightness & According to the 2MASS \\
6 & Flux & Observed bolometric & Integral under interpolated\\
  &      &  flux, W/m$^2$ &  SED \\
7 & Luminosity & Object luminosity & At the Distance with account of\\
  &     &       &  interstellar absorption \\
8 & Distance & Distance in parsecs & In the absence of distance data,  \\
  &          &                 &   $nan$ is put and the luminosity is assumed \\
  &          &                 &   for 1000~pc and $E(g-r)=0$ \\
9 & $E(g-r)$ & Color excess & $nan$, if unknown\\
10 & Phot Pic & SED graph from observational data & PNG-file  \\
11 & SED Pic & SED graph with data interpolation & \parbox{200pt}{PNG-file. The symbol ''+'' means that the ISO SWS spectrum matches the photometric data; the symbol ''-'' means that there is a significant discrepancy between the spectral and photometric data; the symbols ''+-'' mean that there is a good match between the spectral and photometric data over a significant wavelength interval.} \\
12 & Phot dat & SED from observational data & TXT-file \\
13 & SED dat & Interpolated SED & TXT-file \\
14 & JSON dat & All data in JSON format & JSON object \\
15 & Archive & All data and ISO spectra & ZIP-file \\
\hline
\end{tabular}
\label{table:tabl1}
\end{center}
\end{table}

Each object in the catalog corresponds to a record containing 15 fields. These are standard fields with star name, coordinates, and brightness, and fields with compiled and interpolated SED (in the form of text tables, figures, and machine-readable data), as well as an archive with all data and ISO spectra. A more detailed description of the catalog entries is given in Table ~\ref{table:tabl1}.

\begin{figure}[h]
\center{\includegraphics[width=0.9\linewidth]{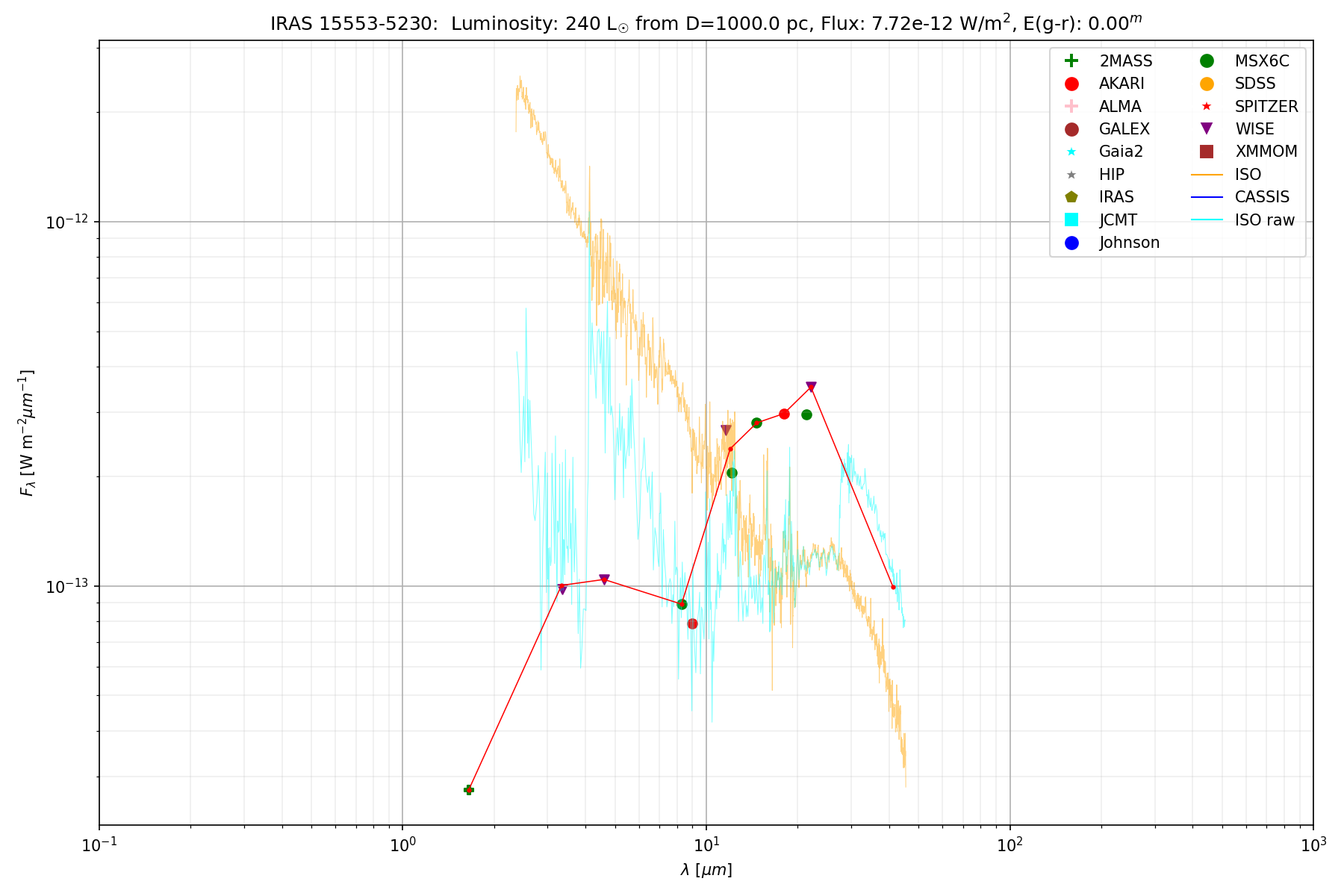}}
\caption{SED of the post-AGB object IRAS 15553-5230. The designations are the same as in Fig.~\ref{fig:fig1}}
\label{fig:fig3}
\end{figure}

In the "SED pic"{} field, in addition to the link to the plot with compiled SED points, ISO spectrum and interpolated SED, there is an indication of the degree of agreement of the ISO SWS spectrum from the Atlas with the general SED curve plotted from the photometric data. The ''+'' symbol next to the link indicates a good match between the ISO spectrum and the photometric data. Such spectra can be automatically combined with photometric data into a common SED in a wide spectral range and used, for example, in modeling of circumstellar dust envelopes. The symbols ''+\,-'' next to the graph link mean that the ISO spectrum partially corresponds to the SED and in this part can be used without recalibration. If there is a ''-'' symbol next to the SED Pic plot link, it is practically impossible to use ISO spectra for this object without recalibration. An example of such a case is shown in Fig.~\ref{fig:fig3}. It clearly shows that the shape of the ISO spectrum of the post-AGB star IRAS~15553-5230 differs from the photometric flux estimates over the entire wavelength range~--- the photometric SED has a double-hump shape and belongs to a cold object with $T_{\rm eff}\sim 1000$\,K with a dust shell, while in the ISO spectrum from Atlas we see a flux drop typical of the Rayleigh-Jeans region of an object with a temperature >3000\,K.

\section*{Carbon stars with dust shells}
\mbox{}\vspace{-\baselineskip}

Using the compiled catalog, we made a selection of carbon stars for future studies of their circumstellar dust envelopes, as we did for V~CrB in the \cite{Fedoteva2020} and for T~Dra in the \cite{Tatarnikov2024}. In addition to the type of star, the main selection criteria were 1) the coincidence of the shape of the photometric SED and the ISO spectrum (i.e., the presence of the "+"{} sign in the Sed Pic column) and 2) the culmination at the air mass $M_z<2$ at the CMO and the Crimean Astronomical Station of SAI. To improve the accuracy of the luminosity determination, we chose the distance limit~--- $D<1500$~pc as an additional selection criterion. Twenty-seven catalog objects satisfy these criteria: AFGL~2699, HV~Cas, IRC~+00365, IRC~+10216, IRC~-10095, PQ~Cep, RU~Vir, RY~Dra, S~Cep, S~Sct, SS~Vir, T~Dra, TT~Cyg, TU~Tau, TX~Psc, U~Cam, V~Aql, V~CrB, V~Cyg, V460~Cyg, V623~Cas, V636~Mon, V833~Her, VX~And, W~Cas, W~Ori, and Y~CVn.

\begin{table}[t]
\caption{IR photometry of selected catalog objects}
\begin{center}
\begin{tabular}{| l | l | l | l | l |}
\hline
Object & Date  & $K$ & $L$ & $M$ \\
\hline
CW Leo & 2024-01-29 & $-0.22_{\pm 0.03}$ & $-3.75_{\pm 0.01}$ & $-4.83_{\pm 0.05}$ \\
HV Cas & 2024-01-30 & $2.21_{\pm0.05}$   & $0.69_{\pm 0.05}$  & $0.66_{\pm 0.05}$  \\
RU Vir & 2024-01-25 & $1.73_{\pm0.02}$   &                    & $-0.12_{\pm 0.05}$ \\
RW LMi & 2024-01-29 & $0.61_{\pm0.02}$   & $-2.21_{\pm 0.02}$ & $-2.86_{\pm 0.07}$ \\
RX Boo & 2024-01-28 & $-1.88_{\pm0.02}$  & $-2.42_{\pm 0.05}$ & $-2.15_{\pm 0.05}$ \\
RY Dra & 2024-01-30 & $0.16_{\pm0.02}$   & $-0.62_{\pm 0.02}$ & $0.12_{\pm 0.02}$  \\
S Cep  & 2024-01-25 & $-0.15_{\pm0.05}$  &                    & $-1.39_{\pm 0.02}$ \\
T Lyn  & 2024-01-30 & $2.88_{\pm0.05}$   & $1.92_{\pm 0.03}$  & $1.80_{\pm 0.03}$  \\
U Cam  & 2024-01-30 & $0.37_{\pm0.05}$   & $-0.36_{\pm 0.05}$ & $0.41_{\pm 0.10}$  \\
V623 Cas & 2024-01-30 & $1.15_{\pm0.05}$   & $0.46_{\pm 0.02}$  & $0.94_{\pm 0.11}$  \\
W Ori  & 2024-01-30 & $-0.44_{\pm0.03}$   & $-1.07_{\pm 0.03}$  & $-0.74_{\pm 0.04}$  \\
\hline
\end{tabular}
\label{table:phot}
\end{center}
\end{table}

The first results of photometric observations of a sample of carbon stars with the LMP IR camera are presented in Table ~\ref{table:phot}. The photometric accuracy was checked by measurements of several standards located at similar air masses. It can be seen that in the mid-IR range all these stars are bright sources whose photometric accuracy is determined not by noise (photon or instrumental) but by the photometric reference to standards located in other parts of the sky. The $K-L$ color indices of all objects are much larger than those of normal stars (see, e.g., \cite{Koornneef}), indicating the presence of excess IR radiation associated with circumstellar dust shells.

\section*{Conclusions}
\mbox{}\vspace{-\baselineskip}

When modeling the SED of objects with dust shells, it is important to have data in the widest possible spectral range~--- from the UV (especially in the case of hot post-AGB or PPN objects) to the far-infrared. For a reasoned selection of the chemical composition of dust particles, the spectra obtained in the infrared play an important role, primarily in the 8-25~$\mu$m region, where the emission features of silicate dusts, silicon carbide dusts, etc, fall. The near-IR is important in selecting the parameters of the star inside the envelope, especially in the case of cold stars. For example, a comparison of the spectra of carbon stars obtained in the 2~-- 5~$\mu$m region with models of atmospheres from the \cite{Aringer2009} allowed the \cite{Fedoteva2020} and \cite{Tatarnikov2024} to obtain the $T_{\rm eff}$ of a star inside a dense dust shell, and to use a realistic approximation of the star's spectrum in modeling instead of the usually used black-body radiation.

Spectra in a wide range from 2.36 to 45~$\mu$m for a large number of AGB and post-AGB objects were obtained by the Infrared Sspace Observatory. Based on these spectra, presented in the atlas ''An atlas of fully processed spectra from the SWS'' and described in \cite{Sloan2003}, we compiled a catalog of SEDs of the Late Stages Stars (available at \textit{https://infra.sai.msu.ru/sai\_lss\_sed}). In it, SEDs in the range 0.4~-- >100~$\mu$m are presented for 263 objects, bolometric fluxes are calculated, and smoothed SEDs (without and with interstellar absorption) are given. Data from the catalog can be directly used in modeling circumstellar dust envelopes (as, for example, was done in the paper \cite{Tatarnikov2024}~---we used the SED of the carbon star T~Dra from our catalog and additional data from IR spectral and photometric observations).

An important result of the analysis of the catalog data is the estimation of the degree of correspondence of the ISO SWS spectrum to the SED derived from photometric data. It turned out that in the rather popular atlas of Sloan et al. {\cite{Sloan2003} (more than 200 citations by the end of 2023~), only 60\% of objects have ISO SWS spectra corresponding to SED, and can be used as they are given in the atlas. Such objects are marked with ''+'' in our catalog. For $\approx$20\% of objects, a complete recalibration of the raw ISO spectra is required (sign ''-'' in the catalog). The spectra of another 20\% of stars can be used partially, in the region of the spectrum (usually~--- in the long-wavelength region) where they correspond to the SED (sign ''+\,-'' in the catalog).

For bright objects in the catalog, one should pay attention to the WISE photometry data in the $W1$ and $W2$ bands. At fluxes $>5 \cdot 10^{-12}$~W/m$^2$\,$\mu$m the detector overexposure is possible. An indirect sign of overexposure is significant (tenths of magnitude) photometry errors in these bands.

The luminosities of all objects in the catalog are derived from the bolometric flux $F_{\rm bol}$, measured from the smoothed and corrected for interstellar absorption SED, and distance: $L=4\pi D^2 F_{\rm bol}$. Thus, the luminosity strongly depends on the adopted distance to the object, especially for objects located in regions of high interstellar absorption. Despite the use of the most modern distance from the Gaia EDR3 catalog \cite{gaia_edr3}, apparently, for red variable stars of large radius, having extended cold atmospheres and often circumstellar dust shells, the distance errors remain large.

We selected 27 carbon stars to study their circumstellar dust envelopes using the RADMC-3D code for solving the equations of radiative transfer in a dusty medium. To complement SED in the wavelength region 2-5~$\mu$m (including replacing WISE observations of bright sources), observations of these objects with the new LMP infrared camera of the 2.5-m CMO telescope have been started. The obtained data demonstrate the high accuracy of photometry.

\bigskip

The work was performed using equipment purchased under the Development Program of M.~V.~Lomonosov Moscow State University (Scientific and Educational School <<Fundamental and Applied Space Research>>).  The work of  A.~Tatarnikov (problem formulation, data analysis) was supported by the Russian Science Foundation (project 23-22-00182). S.~Zheltoukhov (observations with LMP photometer, calibration and processing of obtained data) acknowledges the support of the Foundation for the Development of Theoretical Physics and Mathematics BASIS (project no. 21-2-10-35-1) . This publication makes use of data products from the WISE, which is a joint project of the University of California, Los Angeles, and the Jet Propulsion Laboratory/California Institute of Technology, funded by NASA. Based on observations with ISO, an ESA project with instruments funded by ESA Member States (especially the PI countries: France, Germany, the Netherlands and the United Kingdom) and with the participation of ISAS and NASA


\end {document}